# Arrival Directions of Ultrahigh Energy Showers

V.A. Kolosov, A.A. Mikhailov
*Yu.G. Shafer Institute of Cosmophysical Resarch and Aeronomy of SB RAS, Yakutsk, Russia*
mikhailov@ikfia.ysn.ru

**Abstract:** Arrival directions of ultrahigh energy extensive air showers (EAS) by world arrays data are analyzed. It is shown that arrival directions of ultrahigh energy EAS by Yakutsk and P. Auger data are some correlated with pulsars of Local Arm Orion of Galaxy. It is found large amplitude of first harmony of Fourier row by data Yakutsk array. The problem of cosmic rays origin is discussed.

## Introduction

We have analyzed arrival directions extensive air showers (EAS) ultrahigh energy by Yakutsk, P. Auger, AGASA arrays data. We consider in a case of Yakutsk data array 2 variants: 1) an arrival directions of EAS with usual content of muon particles, 2) an arrival directions of EAS with deficit content of muon particles.
We have analyzed the Yakutsk EAS array data whose shower cores lie inside the array perimeter and accuracy of arrival angle determination is ~ 3°. The particle energy is estimated with the accuracy ~30%. Three showers have energy E>$10^{20}$ eV. The particle energy by P. Auger data is determined with the accuracy ~22%, the solid angle - ≤1° [1]. The particle energy by data AGASA is determined with the accuracy ~25%, the solid angle - ≤1.6° [2].
We do not analyzed EAS of data HiRes array [3] because this array operate only during dark time of day and without detailed knowledge of registration EAS it is impossible to define exposition from separate parts of celestial sphere to array.

## Experimental Data

Fig.1 presents the distribution of 34 particles by Yakutsk array data with the energy E>$4.10^{19}$ eV on the map of equal exposition of the celestial sphere for the period of 1974-2009 (the method to construct this map is based on the estimation of the expected number of showers in the celestial sphere [4]). On the map of equal exposition the equal number of particles from the equal parts of sphere is expected. The Fig. 1 shows the distribution of the showers in the map of celestial sphere in second system on equatorial coordinates δ (a declination) and RA (a right ascension). As it is seen the distribution of showers is almost isotopic. But 13 EAS are located within the angular cone ϑ<45° from axis b=0°, l=90° (b, l - galactic coordinates) of the axes of Local Arm of the Orion Galaxy. From them 11 EAS are located <6° from pulsars [5].

Fig.2 presents the distribution of 26 particles by Yakutsk array data with deficit muons at the energy $E>8 \cdot 10^{18}$ eV. Distribution of EAS with deficit muons is not isotropic, from a galactic plane some excess of an observed number of EAS is observed: $n(|b|<30°)/n(|b|>30°) = 1.9±0.7$. In case of isotropy this ratio will be equal 1.2 according to [6].

We have found among these EAS 5 doublets and from them 4 doublets are located at one region of a celestial sphere: $\delta=20°-75°$ and RA=60°- 80°. The fifth doublet which consists of two EAS: one without muons and the other - with poor muons, is located near Input of the Local arm of the Orion Galaxy.

From direction RA=60°- 80° we find a maxima distribution EAS with usual content muons at energy interval $E=5 \cdot 10^{18}- 4 \cdot 10^{19}$ eV [7].

Further, we determined a correlation between the arrival directions of EAS with $E>4 \cdot 10^{19}$ eV by world data and pulsars. To this end we took the following directions: (i) over the entire celestial sphere region visible by the array and (ii) along the field lines of large-scale regular magnetic field of Orion Arm within a cone with angles <45º from the field-line axis with the galactic coordinates b=0° and l=±90°. This direction along lines of a magnetic field of Orion was chosen, because the magnetic field minimally deviates particles moving along field lines, the probability of correlation between shower arrival directions and pulsars will be increased.

We calculated the angular distances $\vartheta$ between the arrival direction of each pulsars and all particles. The probability P of chance of the number of particles $N(\vartheta)$ was calculated by the Monte-Carlo method through simulation (this number of particles we determine from experimental data) distributed isotropy within a considered region with take account exposure of celestial sphere to array according [4]. The number of simulation was determined by the accuracy of determining the chance probability that chance probability and reached $10^6$ in some case.

We did not find any correlation between arrival directions EAS of Yakutsk (with usual and deficit muons), AGASA data at energy $E>4 \cdot 10^{19}$ eV with pulsars. But we find some correlation between arrival directions by data EAS P. Auger observatory and pulsars at $E>4 \cdot 10^{19}$ eV inside a solid angle <45º from the axis b=0º and l=270º of Orion Arm (Table 1).

Also we search a correlation between arrival directions by data EAS Yakutsk data with EAS usual muons and pulsars (Table 1) below energy $E>(0.8-4) \times 10^{19}$ eV. We find some correlation between arrival directions and pulsars at angular distance <3°, 6° (Table 1). These results confirm our early result [8].

We consider pulsars around which have number of particles ≥10 at angular distances R<6° by Yakutsk data at $E=(0.8-4) \times 10^{19}$ eV (Fig.3). Majority of these 52 pulsars is situated near a galactic plane.

Number of particles around these pulsars at R<6° is 260 (expected number of particles according method [6] – 200.7), at R<3° - 110 (expected number of particles – 75.8). Chance probabilities of number of particles <R from these pulsars at both cases are $P<10^{-4}$ according [6].

We find for these energy $(0.8-4) \times 10^{19}$ eV ratio number of particles n at latitudes b – $n(|b|<30°)/n(|b|>30°)=1.4±0.09$ (expected number of particles according [6] – 1.22). Excess number of particles than expected ones at isotropy is ~ $2\sigma$.

It is shown maximum concentration pulsars at galactic coordinates b~0° and l~60°, l~115°, l~130° which give maximum direct particles from a galactic plane (we show this result early [7]).

We divided an observed region of energy $>10^{18}$ eV into 4 intervals: 1) $10^{18} – 5 \cdot 10^{18}$ eV, 2) $5 \cdot 10^{18} – 8 \cdot 10^{19}$ eV, 3) $8 \cdot 10^{19} – 4 \cdot 10^{19}$ eV, 4) $>4 \cdot 10^{19}$ eV. Results of harmonic analysis by harmonic functions of Fourier row of the distribution of particles in right ascension are shown in Fig.4. In interval energy $8 \cdot 10^{19} –$

$4 \cdot 10^{19}$ eV the amplitude of 1-st harmonic of distribution in right ascension is $A_1=15.2\pm4.8\%$ and its phase RA=0°. Probability of chance of amplitude is equal P~0.006 according [9]. Number of EAS is n=898.

We found 6 pulsars PSR 1332-3032, 1308-4650, 1355-5153, 1405-5641, 1308-5844, 1314-6101 for P. Auger data [1] from these pulsars it is observed at angular distances at R<6° 12 particles (Fig.5). Chance probability to observe at angular distances R<6° 12 particles from 69 is $P<10^{-4}$. This result confirms our early result [10] by data [11].

## Conclusion

We found that some correlation arrival directions EAS (by Yakutsk data at $E>8 \cdot 10^{18}$ eV and P. Auger data at $E>4 \cdot 10^{19}$ eV) and pulsars, maximum in the distribution of particles from side of Local Arm Orion by data P. Auger at energy $E>4 \cdot 10^{19}$ eV, by Yakutsk data at $E=(0.8-4)\times 10^{18}$ eV. We found large amplitude of 1-st harmonics $A_1$~15% by Yakutsk data array at $E\sim 10^{19}$ eV. Most likely pulsars are sources of cosmic rays ultrahigh energy.

References

[1] The P. Auger Collaboration. Astropart. Phys., 2010, 34, 314-326.
[2] N. Hayashida, K. Honda, N. Inoue et al. Astrophys. Journ., 1999, 522, 225-265.
[3] R.U. Abbasi, T. Abu-Zayyad, M. Allen et al., arXiv: 0804.0382v2.
[4] N.N. Efimov, A.A. Mikhailov, Astrop. Phys., 1994, V.2, 329 – 333.
[5] R. N.Manchester, G. B. Hobbs, A. Teoh, & M. Hobbs, Astroph. J., 2005, 129, 1993-2006.
[6] P.A. Becker, G.S. Bisnovatyi-Kogan, D. Casadei, …, A.A. Mikhailov et al. Frontiers in cosmic ray research, Monograph, New York, Nova Sc. Publ., 2007, ch. 6.160.
[7] A.A. Mikhailov. Nucl. Phys. B. (Proc. Suppl.) 2011, V.212-213, 213-218.
[8] A.A. Mikhailov. Izvestia AN, ser.phys., 1999, V.63, 556-559.
[9] Linsley J. Phis. Rev. Lett., 1975, V.34, 1530-1533.
[10] A.A. Mikhailov, Efremov N.N., Gerasimova N.S et al. Proc. 21-st ECRS, Kosice, 2008, 442-445.
[11] P. Auger Collaboration. Astroph. Phys. 2008, V.29, 188.

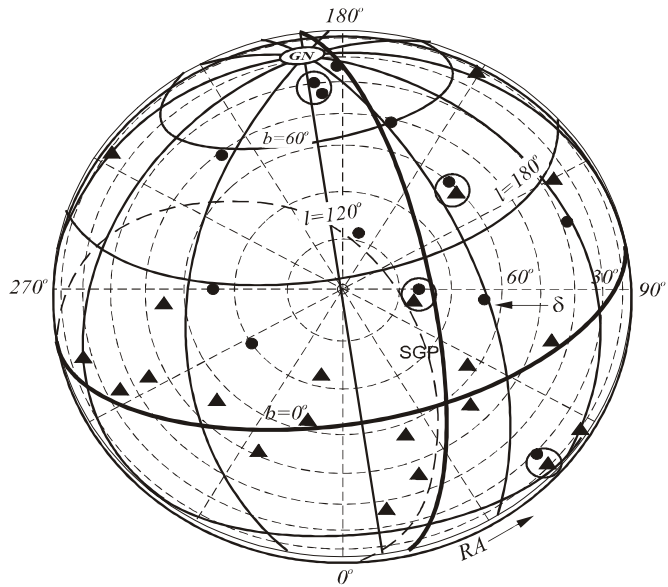

Fig.1. Distribution of EAS with energy E>4.10$^{19}$ eV: δ and RA are declination and right ascension, the dashed curve is the conditional boundary of the Orion Arm, R<45°. ▲, ● - EAS, which correlate and uncorrelated with pulsars accordingly.

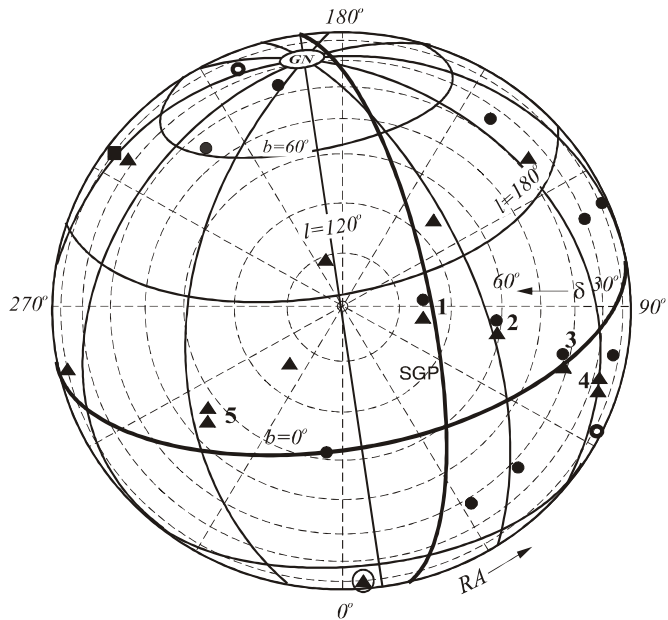

Fig.2. Distribution of EAS with deficit muons: ▲, ● -EAS with without muons, o – EAS with poor muons, other marks are a same as Fig.1. Figures – number of clusters.

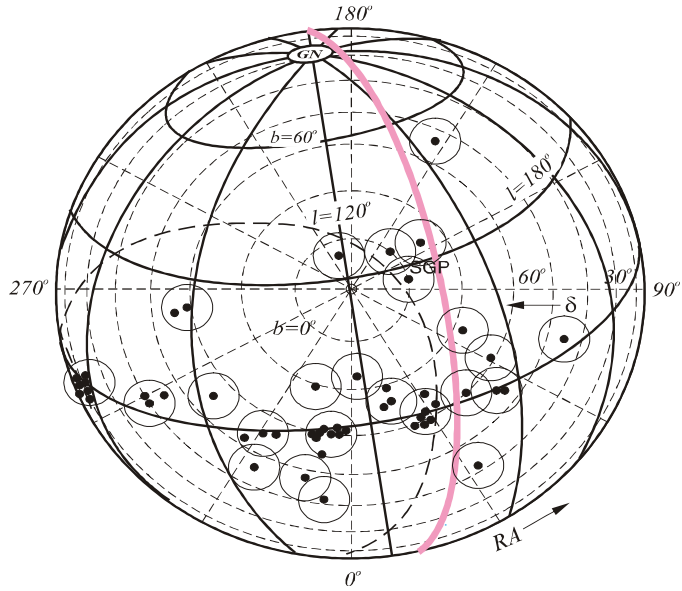

Fig.3. Distribution pulsars which have ≥10 particles with E=(0.8-4)10$^{19}$ eV at radius <6° (circles).

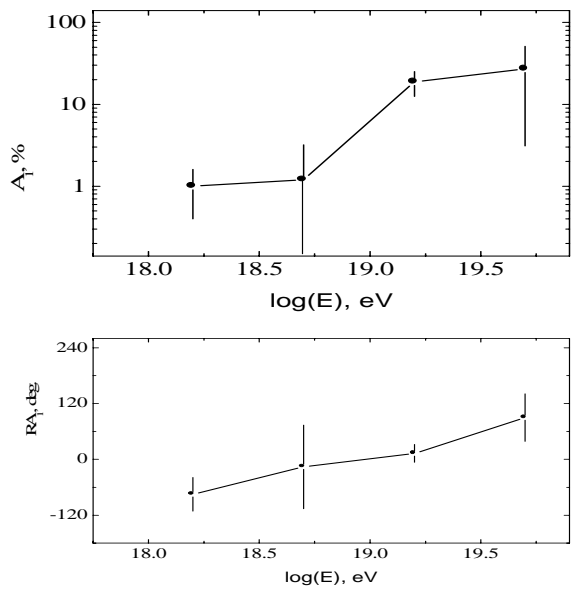

Fig.4. Amplitudes $A_1$ and phases $RA_1$ of 1-st harmonics by Yakutsk data.

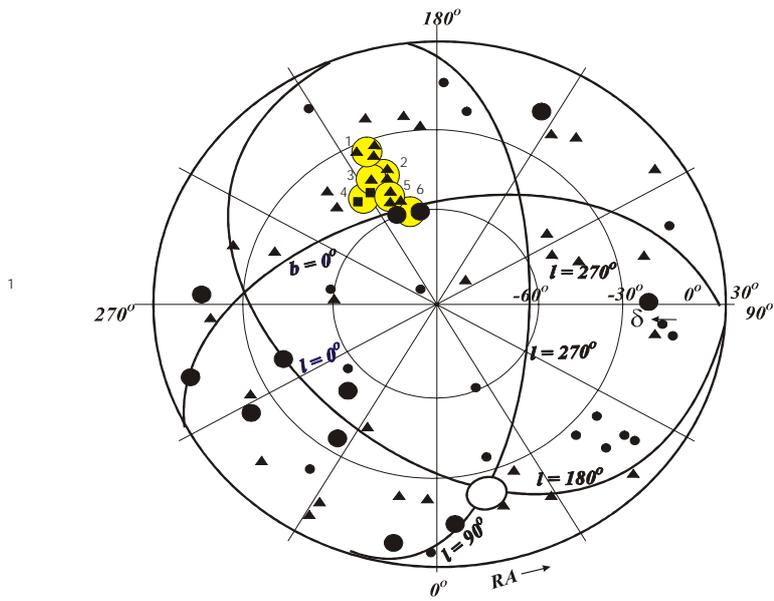

Fig.5. Distribution EAS by by P. Auger data. It is observed maximum in the distribution of particles around 6 pulsars (circles).

Table 1. Correlation EAS with pulsars inside Orion arm.

| Array | E, eV | $\vartheta$, deg. | $N_1$ | $N_2$ | P |
|---|---|---|---|---|---|
| P. Auger | >4.10$^{19}$ | 3 | 69 | 6 | 4.10$^{-2}$ |
| Yakutsk | (0.8-4) 0$^{19}$ | 6 | 898 | 265 | 10$^{-3}$ |
| Yakutsk | (0.8-4)10$^{19}$ | 3 | 898 | 140 | 3.10$^{-2}$ |

Note: $\vartheta$ - $N_1$ – number of considered EAS, $N_2$ – number of EAS inside Arm <45° which correlated pulsars, P – probability.